Native point defects and carbon clusters in 4H-SiC: A hybrid functional study

Takuma Kobayashi[1], Kou Harada[1], Yu Kumagai[2], Fumiyasu Oba[1], and Yu-ichiro Matsushita[1]

[1]Laboratory for Materials and Structures, Institute of Innovative Research, Tokyo Institute of Technology, Yokohama 226-8503, Japan

[2]Materials Research Center for Element Strategy, Tokyo Institute of Technology, Yokohama 226-8503, Japan

We report first-principles calculations that clarify the formation energies and charge transition levels of native point defects and carbon clusters in the 4H polytype of silicon carbide (4H-SiC) under a carbon-rich condition. We applied a hybrid functional that reproduces the experimental bandgap of SiC well and offers reliable defect properties. For point defects, we investigated single vacancies, antisites, and interstitials of Si and C on relevant sites. For carbon clusters, we systematically introduced two additional C atoms into the perfect 4H-SiC lattice with and without removing Si atoms and performed structural optimization to identify stable defect configurations. We found that neutral Si antisites are energetically favorable among Si-point defects in a wide range of the Fermi level position around the intrinsic regime, whereas negatively-charged Si vacancies and a positively-charged Si interstitial on a site surrounded by three Si and three C atoms become favorable under $n$- and $p$-type conditions, respectively. For C-point defects, neutral C antisites are favorable under intrinsic and $n$-type conditions, whereas positively-charged C vacancies become favorable under $p$-type conditions. We also found that a di-carbon antisite is more favorable than a C-split interstitial, which is the most stable form of single C interstitials.

# I. INTRODUCTION

Silicon carbide (SiC) has attracted increasing attention as a suitable semiconductor material to realize low-loss and fast power devices, owing to its superior physical properties (e.g., wide bandgap, high critical electric field, high electron saturation velocity, and high thermal conductivity).[1,2] Among the common polytypes (3C, 4H, and 6H), 4H is in particular favorable for power devices because of its high bulk electron mobility (1000 cm$^2$V$^{-1}$s$^{-1}$ ($\perp$ [0001]) and 1200 cm$^2$V$^{-1}$s$^{-1}$ ($\parallel$ [0001]) at room temperature (RT)) and wide bandgap (3.26 eV at RT).[2] For these reasons, the development of fundamental technologies, such as crystal growth,[3-7] selective doping by ion implantation,[8-12] and metal-oxide-semiconductor (MOS) technologies[13-17] proceeded so far mainly focusing on the 4H polytype. In recent years, SiC has also been regarded as an attractive material for quantum computing.[18-22] Similar to the well-known nitrogen-vacancy (NV) center in diamond,[23] the silicon vacancy ($V_{Si}$) in 4H-SiC is reported to act as a single photon source (SPS) operating up to RT, which also enables coherent control of single electron spins.[18] Thus, understanding of defect properties in 4H-SiC is of great importance both for power electronics and quantum computing.

As for theoretical studies of native point defects in SiC, systematical investigation of single vacancies ($V_C$ and $V_{Si}$)[24-25], antisites (Si$_C$ and C$_{Si}$)[24], and interstitials (Si$_i$ and C$_i$)[24-25] was carried out in 3C-SiC using first-principles calculations. Single vacancies ($V_C$ and $V_{Si}$) and antisites (Si$_C$ and C$_{Si}$) have also been investigated in 4H-SiC.[26] However, as these studies rely on the local-density approximation (LDA),[27,28] the bandgap of SiC is underestimated, which leads to uncertainty in the formation energies and charge transition levels of the defects. For studies using the Hyde-Scuseria-Ernzerhof (HSE06) hybrid functional,[29-31] which reproduces the experimental bandgap of SiC well, defect properties for single vacancies ($V_{Si}$ and $V_C$)[32-33] and vacancy pairs ($V_{Si}$-$V_C$)[33] were reported in 3C-[33]) and 4H-SiC[32-33]), and those for single interstitials ($Si_i$ and $C_i$) and vacancy-type defects including a vacancy-antisite pair ($V_C$-$C_{Si}$) were investigated in 3C-SiC.[34]) However, a systematic study of native point defects including antisites and interstitials in 4H-SiC using the HSE06 hybrid functional is still lacking.

Investigation of carbon clusters is also of high importance. With the thermal oxidation of SiC, some portion of carbon atoms is known to diffuse into the bulk SiC region[35-36]) to fill in the carbon vacancy ($V_C$).[37-38]) As for calculations of carbon clusters in SiC, charge transition levels and dissociation energy (i.e., an energy to remove a single carbon atom from a defect) for carbon interstitial clusters in 3C-SiC and for di-interstitials in 4H-SiC were investigated based on the LDA,[25]) and formation energies of carbon clusters in 3C- and 4H-SiC were reported based on the mixed LDA-exact exchange calculation.[39])

In this paper, we provide systematic calculation results regarding the defect energetics in 4H-SiC. We adopted the HSE06 hybrid functional[29-31]) to reproduce the experimental bandgap of SiC. The formation energies and charge transition levels are calculated both for native point defects and carbon clusters. For point defects, we investigated single vacancies ($V_{Si}$ and $V_C$), antisites ($Si_C$ and $C_{Si}$), and interstitials ($Si_i$ and $C_i$) on relevant sites. For carbon clusters, we systematically introduced two additional carbon atoms into the perfect 4H-SiC lattice with and without removing Si atoms and performed structural optimization to identify the stable defect configurations.

**II. METHODS**

**A. Computational details**

All of the calculations in this study were performed using the projector augmented-wave (PAW)

method[40]) as implemented in the Vienna ab initio simulation package (VASP) code.[41,42]) The cutoff energies in the plane wave basis set were set to 550 and 400 eV for the calculations of perfect 4H-SiC and defect systems, respectively. The lattice parameters were optimized in the calculation of perfect 4H-SiC, while those were fixed for defect calculations to the optimized perfect ones. The 4×4×1 4H-SiC supercell composed of 128 atoms was adopted for modelling the defects. The HSE06 hybrid functional[29-31]) was applied in all the calculations. Spin polarization was considered in the defect calculations.

**B. Modeling of defects**

We first optimized the structure of perfect 4H-SiC crystal by relaxing both the lattice constants and the internal atomic positions in the 8-atom primitive cell. During this calculation, 8×8×2 $k$-points were sampled and the geometry optimization was performed until the residual forces were reduced to less than 5 meV/Å. The calculated lattice constants were 3.07 Å and 10.05 Å for the $a$- and $c$-axes, respectively, well consistent with the experimental values of 3.08 Å and 10.08 Å at RT[2]).

As depicted in Fig.1, there are two inequivalent lattice sites (h- and k-sites in Fig.1(b)) in 4H-SiC. We investigated all the single vacancies and antisites at the inequivalent sites. For interstitials, we introduced either Si or C atom into site 1, 2, 3, or 4 in Fig.1(b) to investigate the stable interstitial configurations. For carbon clusters, we investigated antisite-antisite pairs ((h,h), (k,k), and (h,k)), antisite-interstitial pairs ((h,2), (h,3), (h,4), (k,1), (k,2), and (k,4)), and interstitial-interstitial pairs ((1,2), (1,3), (1,4), (2,3), (2,4), (3,4), (2,1'), and (4,3')). Here, the combinations of alphabets (h and k) and numbers (1, 2, 3, 4, 1', and 3') indicate which two sites to introduce C atoms. For example, (h,h) indicates the antisite-antisite pair (($C_{Si}$)$_2$) where the C atoms are introduced in the two neighboring Si h-sites, and (2,1') indicates the di-carbon interstitial defect (($C_i$)$_2$) where the C atoms are introduced in interstitial sites 2 and 1'. In these defect calculations, structural optimization was performed till the residual forces became less than 40 meV/Å.

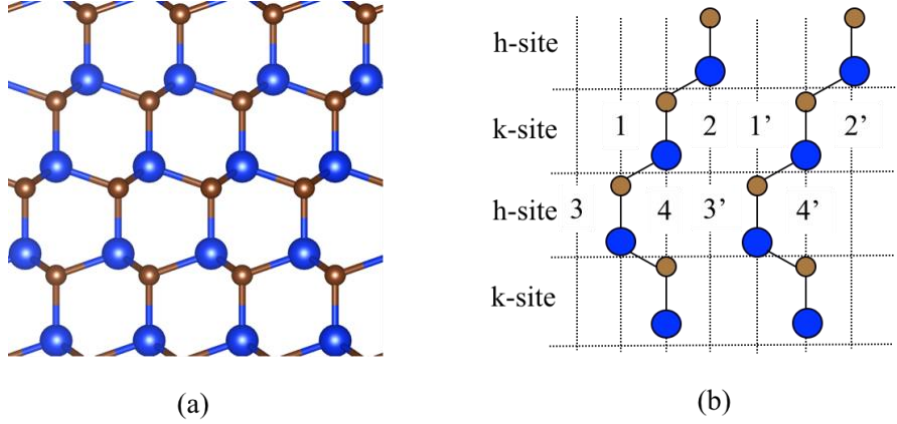

(a)  (b)

Fig. 1: (a) Structure of 4H-SiC lattice viewed from the [11$\bar{2}$0] direction. Blue and brown balls indicate the Si and C atoms, respectively. (b) Schematic structure of the 4H-SiC lattice. There are two inequivalent lattice sites (h- and k-sites) and four interstitial sites (1, 2, 3, and 4). Sites 1', 2', 3', and 4' are crystallographically equivalent to sites 1, 2, 3, and 4, respectively, but used to describe the defect pair combinations.

**C. Defect formation energy**

The formation energy $E_f$ of defect $D$ in charge state $q$ ($D^q$) is calculated as[43-45]

$$E_f[D^q] = (E[D^q] + E_c[D^q]) - E_p - \sum_i n_i \mu_i + q(\varepsilon_{VBM} + \Delta\varepsilon_F), \quad (1)$$

where $E[D^q]$ and $E_p$ are the total energies of the supercell with and without defect $D^q$, respectively. $n_i$ is the number of the removed ($n_i < 0$) or added ($n_i > 0$) $i$-type atom and $\mu_i$ is its chemical potential. $\Delta\varepsilon_F$ is the Fermi level with respect to the energy level of VBM, $\varepsilon_{VBM}$, in perfect 4H-SiC. $E_c[D^q]$ is an image charge correction term associated with spurious electrostatic interactions between $D^q$, its periodic images, and the jellium background under periodic boundary conditions. For the correction term, we applied the extended Freysoldt-Neugebauer-Van de Walle (FNV) scheme, which can correct energies of charged defects accurately in various systems.[43,46-47] Note that, the conduction band minimum (CBM) at the M-point and the valence band maximum (VBM) at the Γ-point in the primitive cell are both folded onto the Γ-point in the 4×4×1 supercell. The calculated bandgap was 3.17 eV, well consistent with the experimental value (3.26 eV at RT).[2]

In investigating the dominant defect species, we first performed Γ point-only calculations with considering the relevant defects described in Sec. IIB and excluded the defects with formation energies higher than 5 eV at any Fermi level position locating inside the bandgap from the candidate defects. At this stage, all the antisite-antisite and interstitial-interstitial C-pairs were excluded. We then increased the number of $k$-points to 2×2×2 and performed geometry optimization for the remaining defects.

In this study, we considered the C-rich limit, in which Si and C chemical potentials are given as

$$\mu_{Si} = E(SiC) - E(C_{diamond}), \qquad (2)$$

$$\mu_C = E(C_{diamond}), \qquad (3)$$

respectively, where $E(SiC)$ and $E(C_{diamond})$ are the total energies of SiC and diamond per formula unit, respectively. Note that the calculated energy difference of diamond and graphite is only about 0.06 eV/atom in the HSE06 calculations. When the Si-rich condition is instead considered, where $\mu_{Si}$ and $\mu_C$ are determined as

$$\mu_{Si} = E(Si_{crystal}), \qquad (4)$$

$$\mu_C = E(SiC) - E(Si_{crystal}), \qquad (5)$$

the change in the $\mu_{Si}$ and $\mu_C$ are only about 0.6 eV, owing to the small formation energy of SiC. Here,

$E(\text{Si}_{\text{crystal}})$ is the total energy of crystal Si per formula unit.

**D. Image charge correction**

As we already mentioned in Sec. IIB, we applied the extended FNV scheme for the correction term $E_c[D^q]$ in Eq. (1).[43, 46-47] To verify that the correction is successful in our systems, we checked whether the formation energy converges with respect to the supercell size by taking a positively-charged carbon vacancy at the k-site ($V_C(k)^{2+}$). In Fig.2, we compare the results for uncorrected, point-charge (PC) corrected[48] and extended FNV corrected[43,46] results. For PC corrected results, the energies are not sufficiently converged with the 128-atom supercell. In contrast, size dependence is dramatically reduced with the extended FNV corrections, and the 72-atom (3×3×1) supercell shows almost the same energy as that of the 128-atom (4×4×1) supercell, indicating that the 128-atom (4×4×1) supercell together with the extended FNV corrections is quite sufficient to obtain well-converged formation energy for charged defects in 4H-SiC.

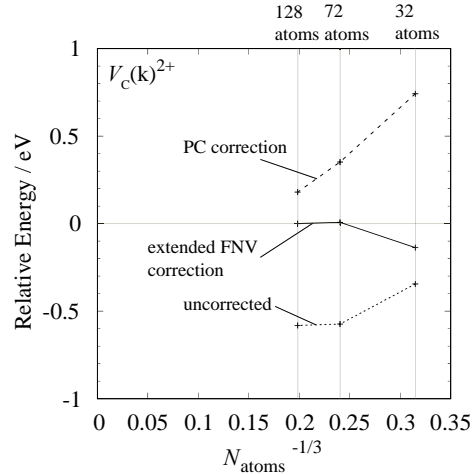

Fig. 2: Relative energy of a carbon vacancy at the cubic (k)-site in the 2+ charge state ($V_C(k)^{2+}$) in 4H-SiC calculated using three supercell sizes (2×2×1 (32 atoms), 3×3×1 (72 atoms), and 4×4×1 (128 atoms)) without any corrections (uncorrected) and with point-charge (PC) and extended Freysoldt-Neugebauer-Van de Walle (FNV) corrections. The horizontal axis is taken as the inverse of the cubic root of the number of atoms in the supercells.

**III. RESULTS AND DISCUSSION**

**A. Native point defects**

Let us first discuss the calculation results for native point defects. Figures 3 (a)-(c) show the structures and formation energies of native point defects in 4H-SiC. We found that neutral Si antisites ($Si_C$) are energetically favorable among Si-point defects ($V_{Si}$, $Si_i$, $Si_C$) in a wide range of the Fermi level position around the intrinsic regime. Negatively-charged Si vacancies ($V_{Si}$) and a positively-charged Si interstitial ($Si_i$(k)) on a site surrounded by three silicon and three carbon atoms also become favorable under *n*- and *p*-type conditions, respectively. For C-point defects ($V_C$, $C_i$, $C_{Si}$), neutral C antisites ($C_{Si}$) are favorable under intrinsic and *n*-type conditions, whereas positively-charged C vacancies ($V_C$) become favorable under *p*-type conditions. Even though C-on-Si antisites are relatively favorable, they do not induce defect levels inside the bandgap of 4H-SiC, and thus seem electrically inactive. The charge transition levels and the formation energies of the neutral defects are summarized in Table. I and II, respectively. In Tables I and II, we see that the transition levels as well as formation energies for $V_{Si}$ and $V_C$ are well consistent with a previous report that also uses the HSE06 functional.[33] The fact that the charge transition levels (Table I) deviate from a report based on the LDA[26] may be attributable to the well-known bandgap underestimation problem of the LDA; however, the neutral formation energies (Table II) of the defects are relatively consistent.

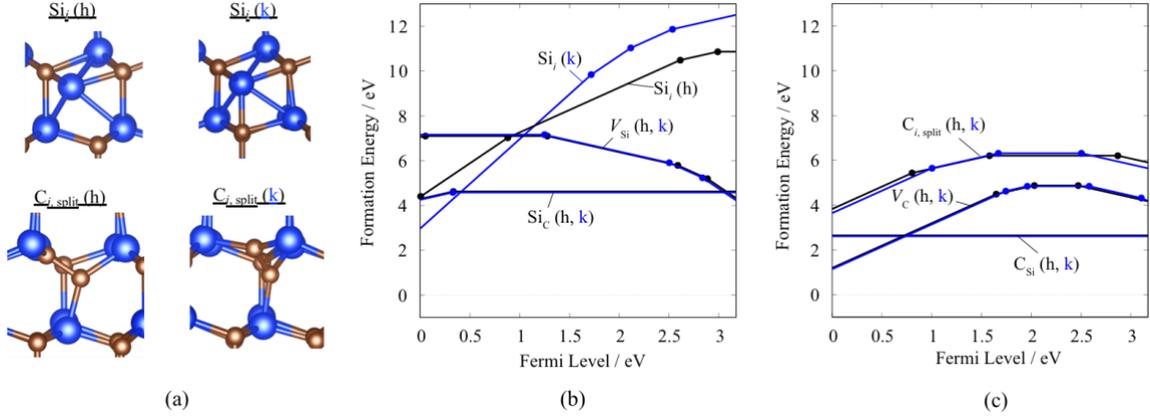

Fig. 3: (a) Optimized structures of neutral Si and C interstitials. The color code is the same as in Fig. 1. (b)(c) Formation energies of the native point defects in 4H-SiC at the C-rich limit; (b) Si-related ($V_{Si}$, $Si_i$, $Si_C$) and (c) C-related defects ($V_C$, $C_i$, $C_{Si}$). The zeros of the Fermi level are set at the VBM and the upper limit corresponds to the CBM.

Table. I: Charge transition levels for the native point defects in 4H-SiC. The levels are given with respect to the energy of the VBM. The calculated values in Refs. 26 and 33 are also shown here for comparison. The data from Ref. 33 are extracted from a figure.

|  | (+4/+3) | (+3/+2) | (+2/+1) | (+1/0) | (0/-1) | (-1/-2) | (-2/-3) | (-3/-4) |
|---|---|---|---|---|---|---|---|---|
| $V_{Si}$ (k) |  |  |  |  | 1.25 | 2.50 | 2.84 |  |
| $V_{Si}$ (h) |  |  |  | 0.05 | 1.28 | 2.59 | 2.89 |  |
| $V_{Si}$[a] |  |  |  | 0.1 | 1.5 | 2.4 |  |  |
| $V_{Si}$ (k)[b] |  |  |  |  | 0.57 | 1.60 | 2.14 | 2.67 |
| $V_{Si}$ (h)[b] |  |  |  | 0.08 | 0.64 | 1.58 | 2.26 | 2.66 |
| $Si_i$ (k) | 1.72 | 2.11 | 2.53 |  |  |  |  |  |
| $Si_i$ (h) | 0.01 | 0.88 | 2.61 | 2.99 |  |  |  |  |
| $Si_C$ (k) |  |  |  | 0.33 |  |  |  |  |
| $Si_C$ (h) |  |  |  | 0.34 |  |  |  |  |
| $Si_C$ (k)[b] |  |  |  | 0.21 |  |  |  |  |
| $Si_C$ (h)[b] |  |  |  | 0.24 |  |  |  |  |
| $V_C$ (k) |  |  | 1.74 | 1.96 | 2.58 | 3.10 |  |  |
| $V_C$ (h) |  |  | 1.65 | 2.03 | 2.47 |  |  |  |
| $V_C$[a] |  |  | 1.7 | 2.0 | 2.5 |  |  |  |
| $V_C$ (k)[b] |  |  | 1.18 | 1.22 | 2.28 | 2.41 |  |  |
| $V_C$ (h)[b] |  |  | 0.97 | 1.34 | 2.09 | 2.21 |  |  |
| $C_{i, split}$ (k) |  |  | 1.00 | 1.67 | 2.50 |  |  |  |
| $C_{i, split}$ (h) |  |  | 0.80 | 1.58 | 2.87 |  |  |  |
| $C_{Si}$ (k) |  |  |  |  |  |  |  |  |
| $C_{Si}$ (h) |  |  |  |  |  |  |  |  |

[a]Gordon et al. (HSE with FNV correction)[33]
[b]Torpo et al. (LDA with Madelung correction)[26]

Table. II. Formation energies for neutral native point defects at the C-rich limit in 4H-SiC. The calculated values in Refs. 26 and 33 are also shown here for comparison. The data from Ref. 33 are extracted from a figure.

| | This work | Ref.[a] | Ref.[b] |
|---|---|---|---|
| $V_{Si}$ (k) | 7.16 | 7.4 | 7.70 |
| $V_{Si}$ (h) | 7.10 | | 7.73 |
| $Si_i$ (k) | 12.46 | | |
| $Si_i$ (h) | 10.87 | | |
| $Si_C$ (k) | 4.58 | | 4.95 |
| $Si_C$ (h) | 4.63 | | 4.92 |
| $V_C$ (k) | 4.84 | 5.2 | 4.41 |
| $V_C$ (h) | 4.88 | | 4.49 |
| $C_{i,\,split}$ (k) | 6.32 | | |
| $C_{i,\,split}$ (h) | 6.21 | | |
| $C_{Si}$ (k) | 2.61 | | 2.85 |
| $C_{Si}$ (h) | 2.65 | | 2.90 |

[a]Gordon et al. (HSE)[33]
[b]Torpo et al. (LDA)[26]

In investigating the point defects, we found several meta-stable structures especially for C-interstitials. Figure 4 shows the structures and formation energies of stable and meta-stable C interstitials. The most stable C interstitials are split-type interstitials ($C_{i,\,split}$), of which splitting direction is not parallel to the c-axis; we found split interstitials parallel to the c-axis ($C_{i,\,split\,(//c)}$) is a meta-stable form. Although a non-split-type interstitials ($C_i$) also appeared as a local minimum structure, its formation energy is quite high compared to the split-type ones and thus should be hardly detected by experiment.

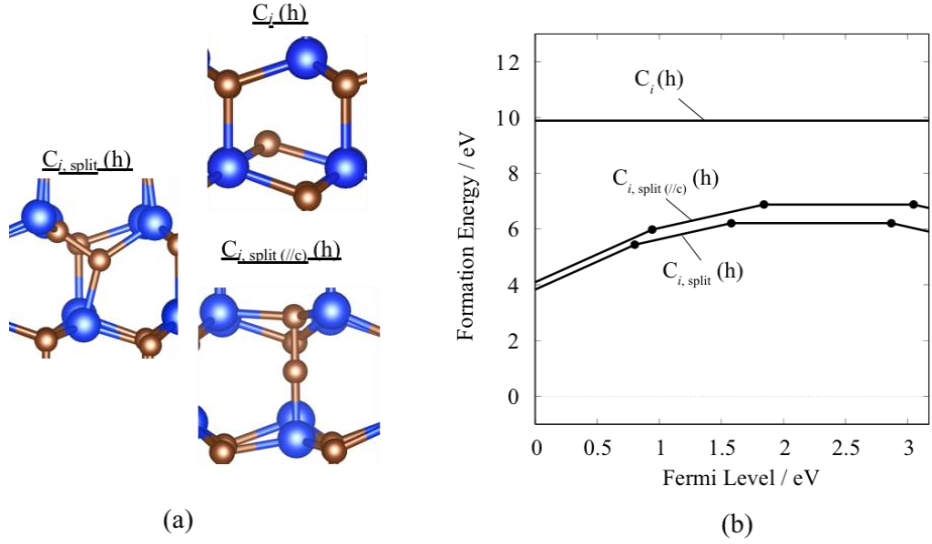

Fig. 4: (a) Optimized structures of stable and meta-stable C interstitials in the neutral charge states in 4H-SiC. The color code is the same as in Fig. 1. (b) Comparison of formation energies for stable and meta-stable C interstitials at the C-rich limit. The zero of the Fermi level is set at the VBM and the upper limit corresponds to the CBM.

**B. Carbon clusters**

Here, we discuss the calculation results for the carbon clusters. As we mentioned in Sec. IIB, the antisite-antisite pairs and interstitial-interstitial pairs were excluded because of their higher formation energies. As a result of geometry optimization of antisite-interstitial pairs, we found that the di-carbon antisites (($C_2$)$_{Si}$) are most stable carbon clusters among the investigated configurations. Figure 5 shows the optimized structure of ($C_2$)$_{Si}$ and their formation energies in comparison with those of single split interstitials ($C_{i,\,split}$). Even though $C_{i,\,split}$ is the most stable form for single C-interstitials (see discussion in Sec. IIIA), we see that ($C_2$)$_{Si}$ is even more favorable under the C-rich condition. The charge transition levels of the ($C_2$)$_{Si}$ defects are summarized in Table III.

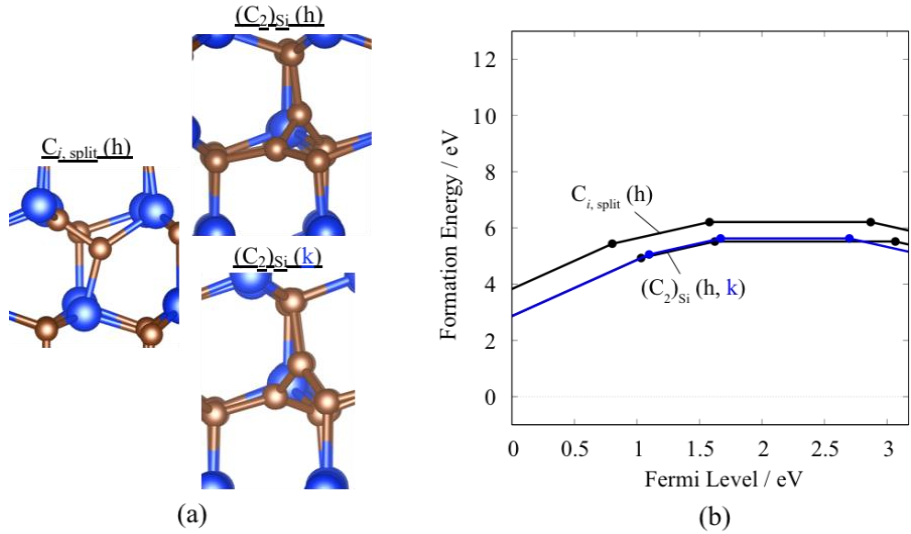

Fig. 5: (a) Optimized structures of neutral single split interstitial and di-carbon antisites in 4H-SiC. The color code is the same as in Fig. 1. (b) Formation energies of the single split interstitial and di-carbon antisites at the C-rich limit. The zero of the Fermi level is set at the VBM and the upper limit corresponds to the CBM.

Table. III: Charge transition levels for di-carbon antisites in 4H-SiC. The levels are given with respect to the VBM.

|  | (+2/+1) | (+1/0) | (0/-1) |
| --- | --- | --- | --- |
| $(C_2)_{Si}$ (k) | 1.10 | 1.67 | 2.70 |
| $(C_2)_{Si}$ (h) | 1.03 | 1.62 | 3.07 |

For the di-carbon antisite, we also found a meta-stable form. In Fig.6 (a), we compare the structures of stable ($(C_2)_{Si}$) and meta-stable ($(C_2)_{Si, tilted}$) di-carbon antisites. For $(C_2)_{Si}$, the carbon atoms C1, C2, C3, and C6 are located on the same plane, as well as C1, C2, C4, and C5, while only C1, C2, C4, and C5 are on the same plane for $(C_2)_{Si, tilted}$. As shown in Fig.6 (b), the formation energy of neutral $(C_2)_{Si, tilted}$ is higher by ~ 0.6 eV than that of neutral $(C_2)_{Si}$. The (+1/0) and (0/-1) transition levels of $(C_2)_{Si, tilted}$ are at 2.21 and 2.27 eV, respectively, and these are very close to each other. The di-carbon antisite previously reported based on a mixed LDA-exact exchange calculation seems to be the $(C_2)_{Si, tilted}$, since the reported (+1/0) and (0/-1) levels are relatively close to each other ((+1/0): 1.4 eV and (0/-1): 1.8 eV).[39]

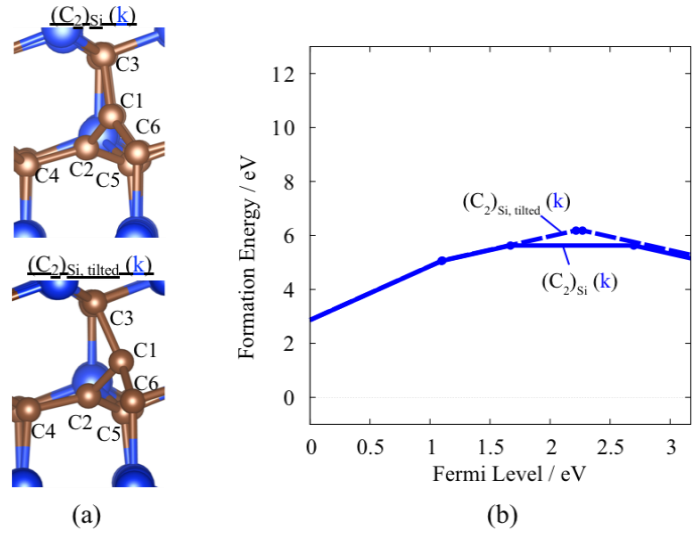

Fig. 6: (a) Optimized structures of stable (($C_2$)$_{Si}$) and meta-stable (($C_2$)$_{Si,\,tilted}$) di-carbon-antisites in the neutral charge states in 4H-SiC. The color code is the same as in Fig. 1. (b) Formation energies of the stable and meta-stable di-carbon antisites at the C-rich limit. The zero of the Fermi level is set at the VBM and the upper limit corresponds to the CBM.

Since we found that the di-carbon antisites (($C_2$)$_{Si}$) are more favorable than single interstitials, we further investigated larger carbon clusters. In 3C-SiC, it was indicated that a tetra-interstitial ring (($C_{sp}$)$_4$ in Ref. 25) owns sizable dissociation energy (5.7 eV), since all carbon atoms are sp$^3$ hybrids.[25] However, we found that the formation energy of the tetra-interstitial ring (see Fig.7 for the relaxed atomic structure) is higher than 13 eV and thus concluded that they are not favorable in 4H-SiC at least when we consider the thermal equilibrium state.

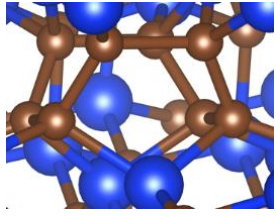

Fig. 7: Optimized structure of the tetra-carbon-interstitial cluster, which is formed by introducing four additional carbon atoms into the 4H-SiC lattice. The structure in the neutral charge state is depicted, and the color code is the same as in Fig. 1.

## IV. SUMMARY

We performed first-principles calculations using the HSE06 hybrid functional to clarify the formation energies and charge transition levels of the native point defects and carbon clusters in 4H-SiC. We found that neutral Si antisites are energetically favorable among Si-point defects in a wide range of the Fermi level position around the intrinsic regime, whereas negatively-charged Si vacancies and a positively-charged Si interstitial on a site surrounded by three Si and three C atoms become favorable under n- and p-type conditions, respectively. For C-point defects, neutral C antisites are favorable under intrinsic and n-type conditions, whereas positively-charged C vacancies become favorable under p-type conditions. By investigating carbon clusters, we found that the antisite-antisite pairs and interstitial-interstitial pairs are unstable, since their formation energies were higher than 5 eV irrespective of the position of the Fermi level. We also found that a di-carbon antisite is more favorable than a C-split interstitial, which is the most stable form of single C interstitials. The formation energy of a tetra-C-interstitial is higher than 13 eV and thus it is not thermodynamically favorable.

**ACKNOWLEDGMENTS**

Computations were performed mainly at the Center for Computational Science, University of Tsukuba, and the Supercomputer Center at the Institute for Solid State Physics, The University of Tokyo. The authors acknowledge the support from JSPS Grant-in-Aid for Scientific Research (A) (Grant Number: 18H03770 and 18H03873).